\documentclass[fleqn,10pt]{CommsPhys} 

\usepackage[english]{babel} 
\usepackage{epsfig}
\usepackage{lipsum} 
\usepackage[super,comma,sort&compress]{natbib}
\usepackage{booktabs}
\usepackage{threeparttable}
\usepackage{caption}
\usepackage[colorlinks,linkcolor=blue]{hyperref}
\usepackage{color}
\def\apj{Astrophys. J.}

\def\aap{Astron. Astrophys.}

\def\nat{Nature}

\def\prd{Phys. Rev. D}
\def\app{Astropart. Phys.}
\def\plb{Phys. Lett. B}
\def\jcap{JCAP}
\def\prl{Phys. Rev. Lett.}
\def\na{Nat. Astron.}
\def\np{Nat. Phys.}

\definecolor{color1}{rgb}{0.0,0.0,.0} 
\definecolor{color2}{rgb}{1.0, 1.0, 1.0} 
\definecolor{color3}{rgb}{1.0,1.0,1.0} 
\definecolor{color4}{rgb}{0.0,0.0,0.0} 
\definecolor{colorbox}{rgb}{0.925, 0.956, 0.992}
\definecolor{colorheader}{rgb}{0.33,0.41,0.47}

\usepackage{hyperref} 
\hypersetup{hidelinks,colorlinks,breaklinks=true,urlcolor=color2,citecolor=color1,linkcolor=color1,bookmarksopen=false,pdftitle={Title},pdfauthor={Author}}

\JournalInfo{Nature Articles} 
\Archive{} 

\PaperTitle{Lorentz violation from gamma-ray burst neutrinos} 

\Authors{Yanqi Huang\textsuperscript{1} and Bo-Qiang Ma\textsuperscript{1,2,3}$^{\star}$} 

\affiliation{\textsuperscript{1}\textit{School of Physics and State Key Laboratory of Nuclear Physics and
Technology, Peking University, Beijing 100871,
China}} 
\affiliation{\textsuperscript{2}\textit{Collaborative Innovation Center of Quantum Matter, Beijing, China}} 
\affiliation{\textsuperscript{3}\textit{Center for High Energy Physics, Peking University, Beijing 100871, China}}
\affiliation{$^{\star}$emall: mabq@pku.edu.cn} 


\title{Lorentz violation from gamma-ray burst neutrinos}

\Abstract{The Lorentz violation~(LV) effect of ultra-relativistic particles can be tested by gamma-ray burst~(GRB) neutrinos and photons. The IceCube Collaboration has observed plenty of ultra-high energy neutrinos, including four events of PeV scale neutrinos. Recent studies suggested a possible energy dependent speed variation of GRB neutrinos in a similar way to that of GRB photons. Here we find that all four events of PeV neutrinos with associated GRB candidates can satisfy a regularity found from TeV neutrinos about a linear form correlation between the observed time difference and the LV factor. Such regularity indicates a Lorentz violation scale $E_{\rm LV}=(6.5\pm 0.4)\times10^{17}~{\rm GeV}$, which is comparable with that determined by GRB photons. We also suggest that neutrinos and anti-neutrinos can be superluminal and subluminal respectively due to opposite signs of LV correction.}

\begin{document}

\fontfamily{lmss}\selectfont
\flushbottom 

\maketitle 


\thispagestyle{empty} 


\rm
\small

\section*{Introduction}

Astrophysical neutrinos are ideal to probe the high energy universe. The IceCube Neutrino Observatory makes it possible to detect and reconstruct ultra-high energy cosmic neutrinos. Based on seven years of measurements, the IceCube Collaboration has observed plenty of neutrino events with energies above 30~TeV, including four PeV scale neutrinos~\cite{Aartsen:2013bka,Aartsen:2013jdh,Aartsen:2014gkd,Kopper:2015vzf,Aartsen:2016ngq}. Probable association between some neutrinos with lower energies and gamma-ray bursts~(GRBs) with close temporal coincidence was suggested by the IceCube Collaboration~\cite{Aartsen:2016qcr,Aartsen:2017wea}. GRB photons and neutrinos may enable us to determine or limit the Lorentz-invariance violation~(LV) physics~\cite{AmelinoCamelia:1997gz,Jacob:2006gn}, since the high energy and the long propagating distance between the GRB source and the detector could produce an observable difference between the GRB trigger time and the arrival time of photons and neutrinos with high energies.
GRB photons with energies above a few tens of GeV were analyzed in previous studies~\cite{Shao:2009bv,Zhang:2014wpb,Xu:2016zxi,Xu:2016zsa,Amelino-Camelia:2016ohi,Xu:2018ien} with the emergence of a remarkable regularity to imply an energy-dependent variation of light speed. Unlike photons, neutrinos are able to escape from dense astrophysical environments and overcome the pair production problem which limits the photon energy. Therefore ultra-high energy cosmic neutrinos provide a powerful tool to explore LV physics~\cite{Jacob:2006gn}. Based on the IceCube data, some research has been done to associate IceCube neutrino events with GRB candidates with longer time range~\cite{Amelino-Camelia:2015nqa,Amelino-Camelia:2016fuh}. The roughly compatible features between GRB photons and neutrinos are revealed by Amelino-Camelia and collaborators~\cite{Amelino-Camelia:2016ohi}. A limit is also proposed from an association between a PeV neutrino event and the outburst of blazar PKS B1424-418~\cite{Kadler:2016ygj,Wang:2016lne}.

In this work, we provide an analysis of the energy dependent speed variation of ultra-high energy IceCube neutrinos, along with the work on TeV scale neutrino events~\cite{Amelino-Camelia:2016ohi}. We find that all four events of PeV scale neutrinos can associate with GRB candidates with much longer time range, and such four events
are consistent with these TeV scale events for a speed variation at $E_{\rm LV}=(6.5\pm 0.4)\times10^{17}~{\rm GeV}$. Lorentz violation also explains the existence of both ``early'' and ``late'' neutrinos.

\section*{Results}
\subsection*{Model.}
For a particle propagating in the quantum spacetime with energy $E\ll E_{\rm Pl}$~(the Planck scale $E_{\rm Pl}\approx1.22\times10^{19}~{\rm GeV}$), the LV modified dispersion relation can be written in a general form as the leading term in Taylor series~\cite{AmelinoCamelia:1997gz,Jacob:2006gn}
\begin{equation}
E^2\simeq p^2c^2+m^2c^4-s_nE^2\left(\frac{E}{E_{{\rm LV}, n}}\right)^n,
\end{equation}
where, $n=1$ or $n=2$ corresponds to linear or quadratic dependence of the energy, $s_n=\pm1$ is the sign factor of LV correction,  $E_{{\rm LV},n}$ is the $n$th-order LV scale to be determined by experiments, and $m$ is the rest mass of the particle. Since photons and ultra-high energy neutrinos are both ultra-relativistic particles, it is reasonable to set $m=0$ in the discussion. Using the relation $v=\partial E/\partial p$, we can get the modified propagation velocity
\begin{equation}
v(E)=c\left[1-s_n\frac{n+1}{2}\left(\frac{E}{E_{{\rm LV},n}}\right)^n\right].
\end{equation}
Such a speed variation can cause a propagation time difference between particles with different energies. By taking into account the cosmological expansion, the LV time correction of two particles with energy $E_{\rm h}$ and $E_{\rm l}$ respectively can be written as~\cite{Jacob:2008bw,Ellis:2002in}
\begin{equation}
\Delta t_{\rm LV}=s_n\frac{1+n}{2H_0}\frac{E_{\rm h}^n-E_{\rm l}^n}{E_{{\rm LV},n}}\int^z_0\frac{(1+z^\prime)^n {\rm d} z^\prime}{\sqrt{\Omega_{\rm m}(1+z^\prime)^3+\Omega_\Lambda}},
\label{time diff}
\end{equation}
where $z$ is the redshift of the GRB source. We adopt the cosmological constants~\cite{Agashe:2014kda} $[\Omega_{\rm m},\Omega_\Lambda]=[0.315^{+0.016}_{-0.017},0.685^{+0.017}_{-0.016}]$ and the Hubble expansion rate $H_0=67.3\pm 1.2~{\rm km\cdot s^{-1}\cdot Mpc^{-1}}$. Here we focus on the $n=1$ case, so equation~(\ref{time diff}) can be rewritten as
\begin{equation}
\Delta t_{\rm LV}=s(1+z)\frac{K}{E_{\rm LV}},
\end{equation}
where $s=\pm1$ is the sign factor and
\begin{equation}
K=\frac{E_{\rm h}-E_{\rm l}}{H_0 }\frac{1}{1+z}\int^z_0\frac{(1+z^\prime) {\rm d} z^\prime}{\sqrt{\Omega_{\rm m}(1+z^\prime)^3+\Omega_\Lambda}},
\label{LV factor}
\end{equation}
is the LV factor. In our discussion, $E_h$ is the neutrino energy~(over 10~TeV) and $E_l$ is the trigger photon energy~(about 100~keV). Since $E_h$ is much higher than $E_l$, $E_l$ in equation~(\ref{LV factor}) is negligible in our analysis.

The observed arrival time difference $\Delta t_{\rm obs}$ between two particles detected on the Earth is actually caused by two reasons, the LV time correction $\Delta t_{\rm LV}$ in the propagation and the intrinsic time difference $\Delta t_{\rm in}$ at the source. $\Delta t_{\rm in}$ is only related to the intrinsic mechanism of the GRB source. Hence we have
 \begin{equation}
 \Delta t_{\rm obs}=t_{\rm h}-t_{\rm l}=\Delta t_{\rm LV}+(1+z)\Delta t_{\rm in}.
 \label{tobs}
 \end{equation}
 where $t_{\rm h}$ and $t_{\rm l}$ represent the arrival times of high-energy and low-energy particles. Considering equation~(\ref{LV factor}), we rewrite equation~(\ref{tobs}) as
\begin{equation}
\frac{\Delta t_{\rm obs}}{1+z}=\Delta  t_{\rm in}+s\frac{K}{E_{\rm LV}}.
\label{linear relation}
\end{equation}
 According to equation~(\ref{linear relation}), there would be a linear relation between $\Delta t_{\rm obs}/(1+z)$ and $K$, if the energy dependence speed variation does exist.

 \begin{table}[t]
 \small
 \centering
 \begin{threeparttable}
      \caption{\small\sf The GRB candidates for the TeV neutrino events.}
      \label{tab:1}
      \centering
      \begin{tabular}{lllll}
       \hline\hline
        \noalign{\vspace{0.5ex}}
         event & GRB & $z$ & $\Delta t_{\rm obs}~(10^3~\rm s)$ & $E$~(TeV)   \\
      \hline
      \noalign{\vspace{0.5ex}}
          $\#2$& 100605A&$1.497^*$&-113.051&117.0   \\

      \noalign{\vspace{0.5ex}}
         $\#9$&  110503A&1.613&80.335&63.2 \\

      \noalign{\vspace{0.5ex}}
         $\#11$& 110531A&$1.497^*$&185.146&88.4  \\

      \noalign{\vspace{0.5ex}}
        $\#12$& 110625B&$1.497^*$&160.909&104.1 \\

      \noalign{\vspace{0.5ex}}
          $\#19$&  111229A&1.3805&73.960&71.5   \\

      \noalign{\vspace{0.5ex}}
          $\#26$& 120219A&$1.497^*$&229.039&210.0  \\

      \noalign{\vspace{0.5ex}}
        $\#33$& 121023A&$0.6^*$&-171.072&384.7  \\

      \noalign{\vspace{0.5ex}}
          $\#40$& 130730A&$1.497^*$&-179.641&157.3   \\

      \noalign{\vspace{0.5ex}}
           $\#42$& 131118A&$1.497^*$&-146.960&76.3   \\

\hline\hline

      \end{tabular}
     \footnotesize
     The nine GRB candidates are suggested from the associated GRBs of IceCube neutrinos with energy between 60~TeV and 500~TeV by the maximum correlation criterion~\cite{Amelino-Camelia:2016fuh,Amelino-Camelia:2016ohi}. The event serial number here is provided by the IceCube database. The mark $^*$ represents the estimated value of the redshift. There are both early neutrinos and late neutrinos.
     \end{threeparttable}
\end{table}

 \subsection*{Analysis of TeV GRB neutrinos.}
 The IceCube Collaboration provided dozens of high-energy neutrino events after seven years of detection~\cite{Aartsen:2013bka,Aartsen:2013jdh,Aartsen:2014gkd,Kopper:2015vzf,Aartsen:2016ngq}. If a neutrino is emitted at the source with an associated GRB, $\Delta t_{\rm obs}$ can be represented by the difference between the arrival time of the neutrino and the trigger time of the given GRB. By using the maximum correlation criterion, Amelino-Camelia and collaborators~\cite{Amelino-Camelia:2016fuh,Amelino-Camelia:2016ohi} selected nine GRB candidates from the associated GRBs of IceCube neutrino events with energy between 60~TeV and 500~TeV. These events with the associated GRBs, the redshift values $z$, the observed time differences $\Delta t_{\rm obs}$ and the neutrino energies $E$ are listed in Table~\ref{tab:1}. The LV factor $K$ of these events can be obtained immediately according to equation~(\ref{LV factor}). The observed time difference $\Delta t_{\rm obs}$ of neutrino events can be positive or negative. $\Delta t_{\rm obs}>0$ events are called ``late neutrinos'', and $\Delta t_{\rm obs}<0$ events are called ``early neutrinos'' in the discussion. However, $\Delta t_{\rm LV}>0 ~(s=1)$ or $\Delta t_{\rm LV}<0 ~(s=-1)$ case is called ``time delay'' or ``time advance'' to avoid confusion. Since $\Delta t_{\rm in}$ is only related to the intrinsic mechanism of the GRB, it is reasonable to expect that $\Delta t_{\rm in}$ is a constant for most events. In fact, $\Delta t_{\rm in}$ can be safely neglected for TeV and PeV neutrinos with $E_{\mathrm{LV}}$ of the scale $10^{18}$~GeV.

 \begin{figure}[t]
\small
\includegraphics[width=8.9cm]{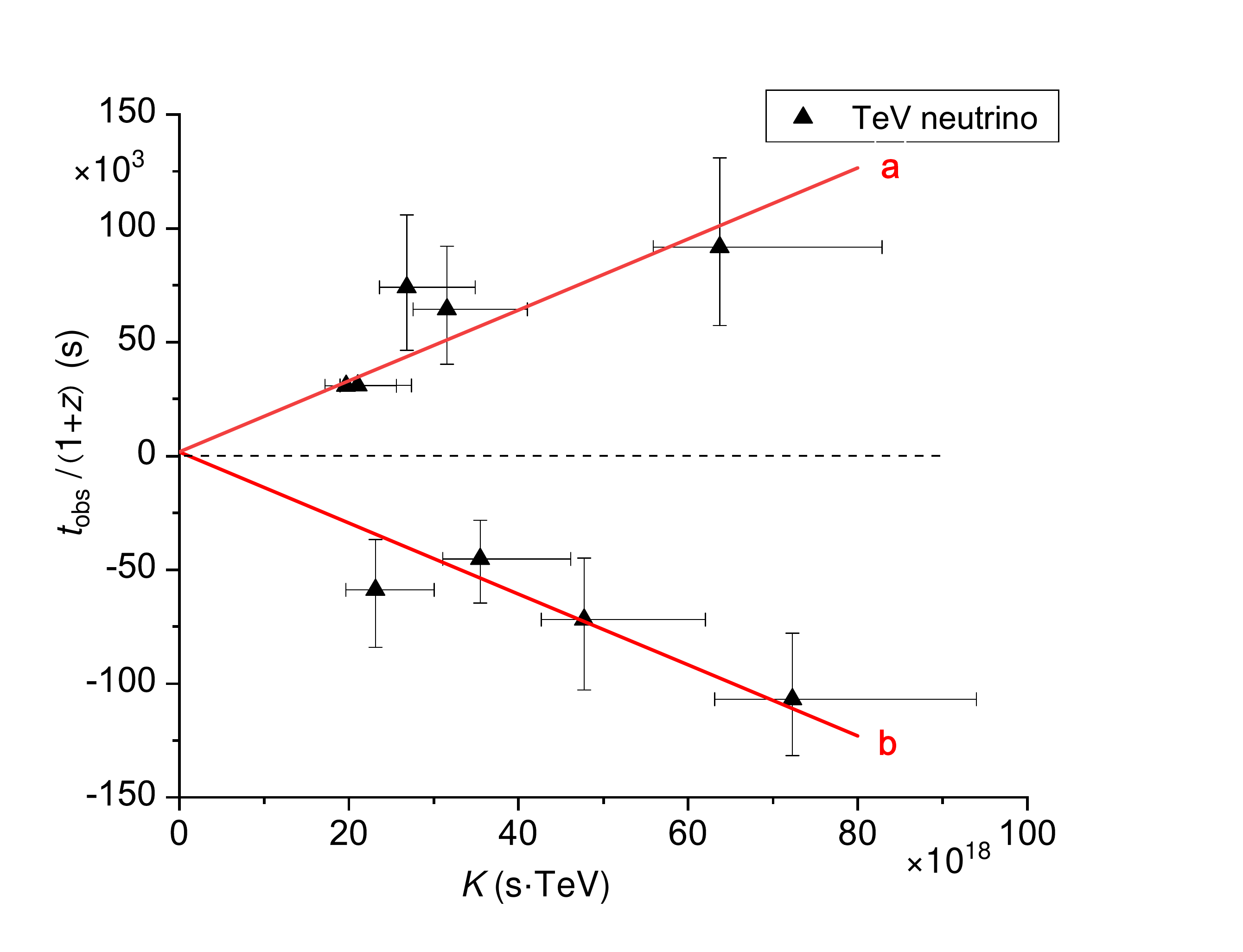}
\caption{\small   \textsf{\textbf{$\Delta t_{\rm obs}/(1+z)$ versus $K$ plot for TeV neutrino events.}}The black triangles are experimentally measured data, and the red lines are linear fits to experimental data. All nine events fall near a pair of lines noted as (a) and (b). The different signs of slopes may be the result of intrinsic distinctions between neutrino and anti-neutrino. The error bars are calculated according to uncertainties of energy, red-shift and cosmological
parameters (see Methods).
}\label{fig1}

\end{figure}

 \begin{figure}[t]
\includegraphics[width=8.9cm]{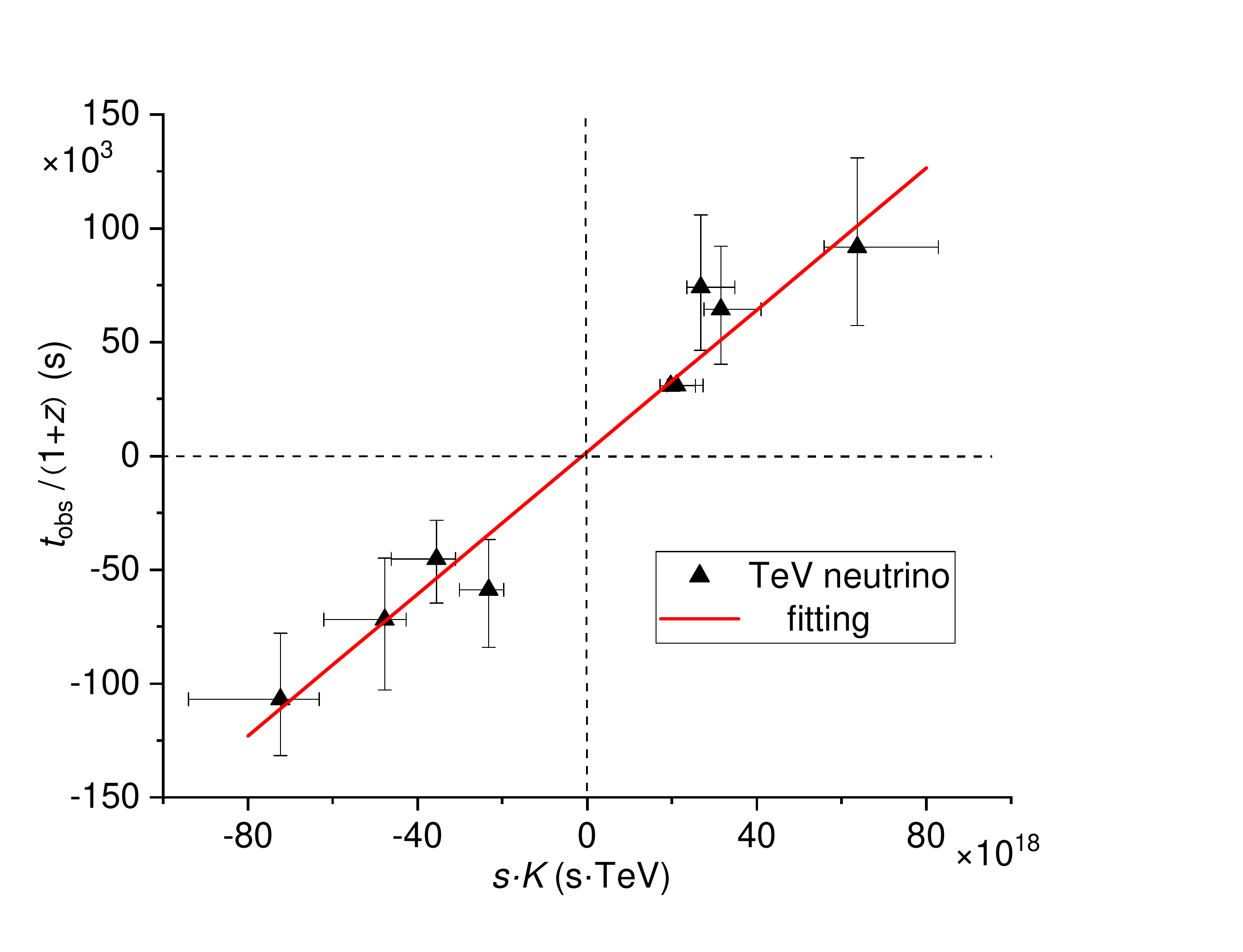}
\caption{\small \textsf{The linear fitting of $\Delta t_{\rm obs}$ and $s\cdot K$ for TeV events.} The black triangles are experimentally measured data, and the red lines are linear fits to experimental data. We set $s=1$ for the five ``time delay'' events, and $s=-1$ for the rest four ``time advance'' events. The slope and intercept are $(1.56\pm0.13)\times10^{-15}~{\rm TeV^{-1}}$ and $(1.8\pm4.1)\times10^3~{\rm s}$. The Pearson's correlation coefficient  $r=0.978$ implies a relatively strong linear correlation between the observed time difference and the LV factor. }\label{fig2}

\end{figure}

To check the possible linear correlation, we draw the $\Delta t_{\rm obs}/(1+z)$ versus $K$ plot for the nine events in Table~\ref{tab:1}, as shown in Fig.~\ref{fig1}. We find that all events fall on a pair of inclined lines (a) and (b) that can be described by the equation
\begin{equation}
|\frac{\Delta t_{\rm obs}}{1+z}-\Delta t_{\rm in}|=\frac{K}{E_{\rm LV}},
\label{abs.}
\end{equation}
which is equivalent to equation~(\ref{linear relation}). The slopes of the pair of lines happen to be opposite numbers, i.e., the LV scales are equal but the sign factors are opposite. Line (a) represents the delay case, and line (b) represents the advance case. It is unnatural that the same kind of particle has two different kinds of propagation properties. Hence, one of the possible interpretations is that the nine events include both neutrinos and anti-neutrinos, which could not be distinguished by the IceCube detector. Since the linear~($n=1$) correction implies the CPT odd term in an effective field theory framework~\cite{Stecker:2014oxa}, neutrinos and anti-neutrinos have different signs
for the LV sign factor $s$. Therefore neutrinos are advanced and anti-neutrinos are delayed, or vice versa.

Taking into account the sign factor $s$, we do a linear fitting for $\Delta t_{\rm obs}$ and $s\times K$ of the events, as shown in Fig.~\ref{fig2}. We set $s=1$ for the five events fallen on line (a), and $s=-1$ for the rest four events fallen on line (b). The slope and intercept are
\begin{eqnarray}
1/E'_{\rm LV}&=&(1.56\pm0.13)\times10^{-15}~{\rm TeV^{-1}},\label{slope}\\
 \Delta t'_{\rm in} &=&(1.8\pm4.1)\times10^3~{\rm s}.\label{intercept}
  \end{eqnarray}
 The Pearson's correlation coefficient $r=0.978$, which implies a relatively strong linear correlation between the observed time difference and the LV factor.
The errors are calculated according to statistical and systematic uncertainties of energy, red-shift and cosmological
parameters (see Methods). Since the error range of $\Delta t_{\rm in}$ covers the zero point, it is still uncertain whether these neutrino events are emitted before or after the GRB photons.

\subsection*{Consistency in PeV neutrinos.}

As mentioned earlier, the influence of the intrinsic time difference and the atmosphere neutrino background would be weakened due to the ultra-high energy of neutrinos. After reanalyzing the nine TeV neutrino events, we turn to the PeV neutrinos. IceCube Collaboration has reported four events with energy $\geq1$~PeV up to now. The events $\#14~(1.04\pm0.16~\rm{PeV})$, $\#20~(1.14\pm0.17~\rm{PeV})$ and $\#35~(2.00\pm0.26~\rm{PeV})$ are based on three years~(2010-2013) of detection~\cite{Aartsen:2013bka,Aartsen:2013jdh,Aartsen:2014gkd}, and the lately reported event ATel $\#7856~(2.6\pm0.3~\rm{PeV})$ is based on an analysis of seven years of data~\cite{Aartsen:2016ngq}. The atmospheric background-only explanation of these PeV events has been rejected at $3.6-5.7~\sigma$~\cite{Aartsen:2013bka,Aartsen:2013jdh,Aartsen:2014gkd,Aartsen:2016ngq}.

To find associated GRBs with neutrino events, we adopt two criteria to restrict the time difference and the direction. At the ultra-high energy scale, a neutrino detected months around the GRB trigger time might be statistically associated with the GRB~\cite{Jacob:2006gn}. So the time range should be expanded with the increase of energy. For each neutrino event, several GRB candidates are selected by the criteria. Their main properties are shown in Table~\ref{tab:2}~(detailed in Methods). The redshifts of some GRBs are not measured yet, here we use the average value of all GRBs observed so far as the ``best guess'' value. According to the GRB database~\cite{IceCube website} provided by the IceCube Collaboration, we set $z=2.15$ for ``long bursts'', and $z=0.5$ for ``short bursts''. In fact, it is not the central value but the error range of the redshift that plays an important role in the analysis~\cite{Amelino-Camelia:2016fuh}. The error estimation of energy and redshift is similar to the former TeV events (see Methods).

\begin{table}[t]
\small
 \centering
\begin{threeparttable}
 \centering
      \caption{\sf \small GRB candidates for the four events of PeV neutrinos. }
      \label{tab:2}

      \begin{tabular}{lllll}
       \hline\hline
        \noalign{\vspace{0.5ex}}
         event & $E$~(PeV)& GRB   & $z$ & $\Delta t_{\rm obs}~(10^3~\rm s)$  \\
      \hline
       \noalign{\vspace{0.5ex}}

       \noalign{\vspace{0.5ex}}
       \    \  &&110725A$^\dagger$& $2.15^*$ & 1320.217 \    \   \\
        \noalign{\vspace{0.5ex}}
       \    \   &&110730A&$2.15^*$ & 907.885 \    \   \\
        \noalign{\vspace{0.5ex}}
       $\#14$ & 1.04 & 110731A & $2.83$ & 782.096 \    \   \\
       \noalign{\vspace{0.5ex}}
       \    \   &&110808B&$0.5^*$ & 74.303\    \   \\
       \noalign{\vspace{0.5ex}}
       \    \   &&110905A&$2.15^*$ & -2309.121\    \   \\
       \noalign{\vspace{0.5ex}}
        \hline

        \noalign{\vspace{0.5ex}}

        \noalign{\vspace{0.5ex}}
        &&111229A &1.3805 & 384.970    \\
       \noalign{\vspace{0.5ex}}
         $\#20$ & 1.14 &120119C$^\dagger$ & $2.15^*$ & -1940.176  \\
       \noalign{\vspace{0.5ex}}
         &&120210A & $0.5^*$ & -3304.901 \\
       \noalign{\vspace{0.5ex}}
        \hline

        \noalign{\vspace{0.5ex}}

        \noalign{\vspace{0.5ex}}
         &&120919A & $2.15^*$ & 6539.722\\
       \noalign{\vspace{0.5ex}}
        $\#35$ & 2.0 &121229A & 2.707 & -2091.621   \\
       \noalign{\vspace{0.5ex}}
         &&130121A$^\dagger$ &$2.15^*$ & -4046.519    \\
       \noalign{\vspace{0.5ex}}
        \hline

        \noalign{\vspace{0.5ex}}

        \noalign{\vspace{0.5ex}}
         ATel & 2.6&140427A$^\dagger$ & $2.15^*$ & 3827.439  \\
       \noalign{\vspace{0.5ex}}
         $\#7856$  &&140516B & $2.15^*$ & 2185.942  \\
        \noalign{\vspace{0.5ex}}

\hline\hline
      \end{tabular}
      \footnotesize
        The GRB candidates here are selected by the time and direction criteria~(detailed in Methods). For every one of the four events, there exists a candidate marked by $^\dag$ that satisfy the strict time criterion and is consistent with the regularity of TeV neutrinos. The mark $^*$ represents a ``best guess'' value of the redshift, i.e., $z=2.15$ for ``long bursts'' and $z=0.5$ for ``short bursts''.
\end{threeparttable}

\end{table}

\begin{figure}[t]
\includegraphics[width=8.9cm]{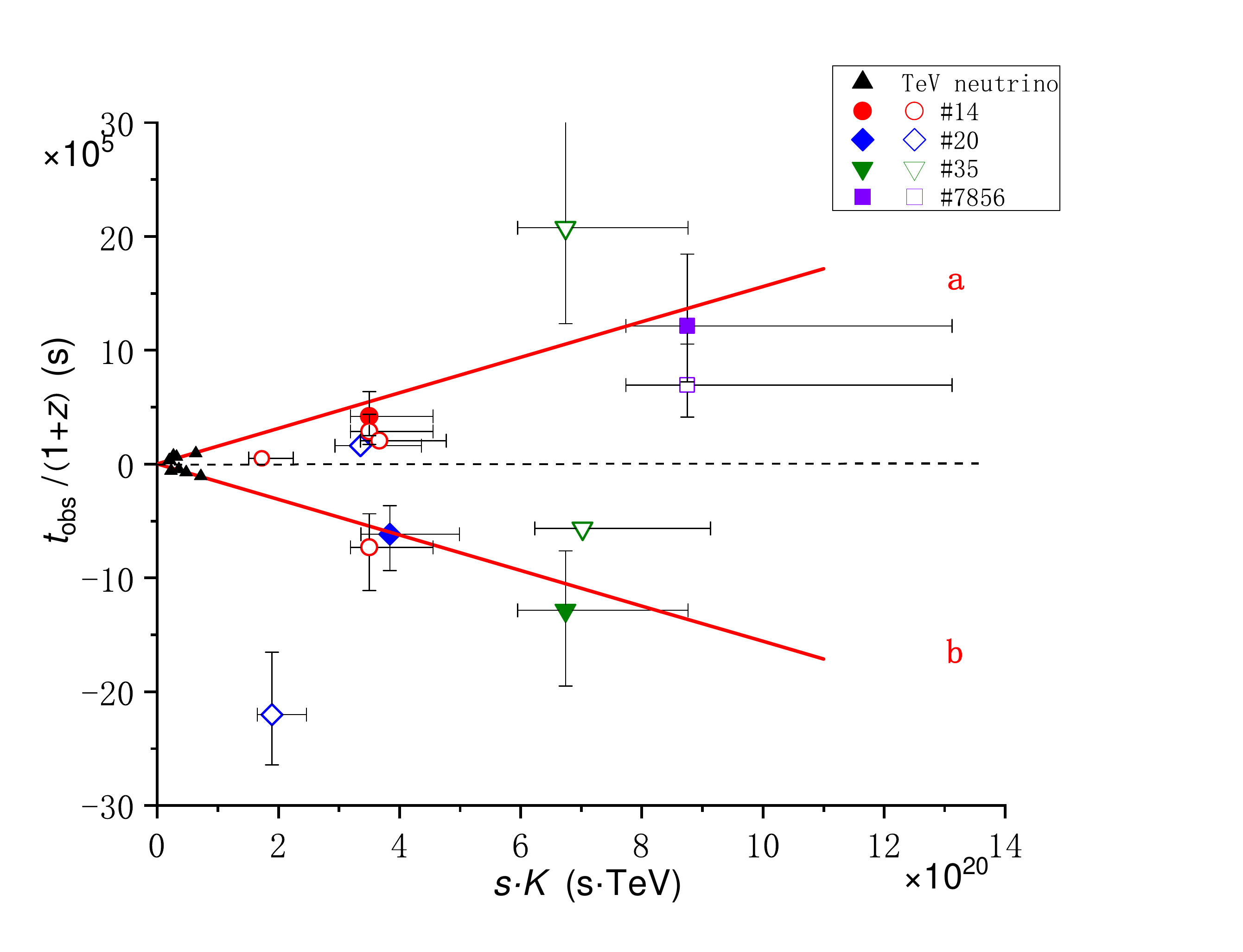}
\caption{\small \textsf{$\Delta t_{\rm obs}/(1+z)$ versus $K$ plot for PeV neutrino events with associated GRB candidates.}  The TeV neutrino events and the pair of lines are drawn here to show the consistency. Different symbols represent the experimental data of different neutrino events, and the red lines are linear fits to TeV neutrino events. The four $^\dag$ marked GRB candidates in Table~\ref{tab:2} are represented as solid symbols and they fall on the same pair of lines (a) and (b) drawn from TeV neutrino events. The open symbols represent the rest candidates that satisfy the direction criterion.
}\label{fig3}

\end{figure}

\begin{figure}[t]
\includegraphics[width=8.9cm]{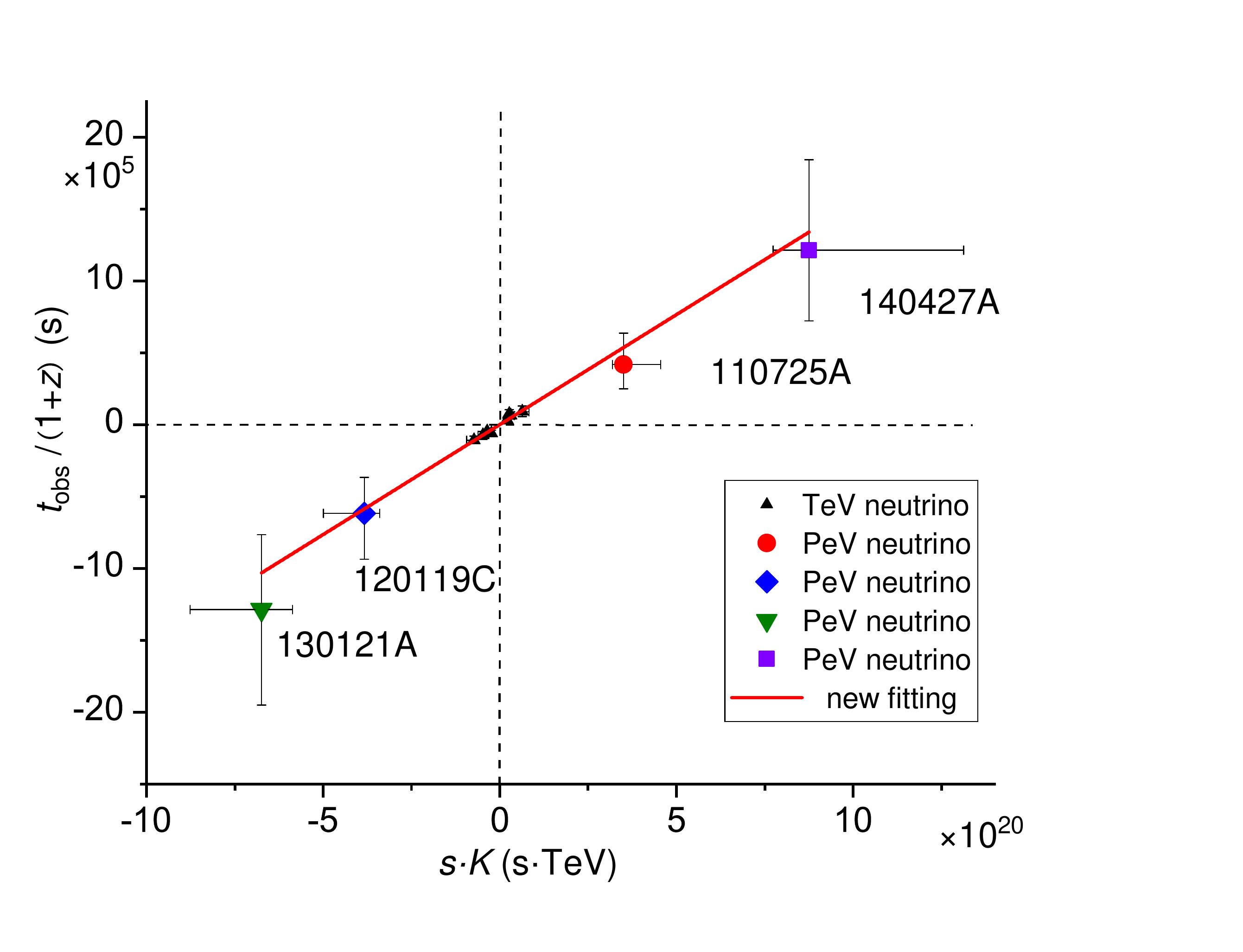}
\caption{\small \textsf{The linear fitting of $\Delta t_{\rm obs}$ and $s\cdot K$ for both TeV and PeV events.} Different symbols represent the experimental data of different neutrino events, and the red lines are linear fits to both TeV and PeV neutrino events. The slope and intercept are $(1.53\pm0.10)\times10^{-15}~{\rm TeV^{-1}}$ and
 $(1.7\pm3.6)\times10^3~{\rm s}$. The Pearson's correlation coefficient  $r=0.98$ also implies a relatively strong linear correlation between the observed time difference and the LV factor. We can obtain the LV scale $E_{\rm LV}=(6.5\pm0.4)\times 10^{17}~\rm{GeV}$, which is comparable with the results proposed by the GRB photon studies~\cite{Zhang:2014wpb,Xu:2016zsa,Xu:2016zxi,Amelino-Camelia:2016ohi,Xu:2018ien}.}\label{fig4}

\end{figure}

By analysing the GRB candidates in Table~\ref{tab:2}, we find that all four events of PeV neutrinos are possible to be in accordance with the regularity of TeV neutrinos on the $\Delta t_{\rm obs}/(1+z)$ versus $K$ plot. Fig.~\ref{fig3} shows the consistency between the TeV candidates suggested by Amelino-Camelia and collaborators~\cite{Amelino-Camelia:2016fuh,Amelino-Camelia:2016ohi} and the PeV candidates suggested by us. The four candidates still fall on the pair of lines, with both time delay and time advance cases. The separated linear fitting of the four events of PeV neutrinos with associated GRB candidates gives a result $1/E''_{\rm LV}=(1.50\pm0.09)\times10^{-15}~{\rm TeV^{-1}}$, which conforms well with equation~(\ref{slope}). Hence, we can do a fitting to all thirteen events, as shown in Fig.~\ref{fig4}. The slope and the intercept are
\begin{eqnarray}
1/E_{\rm LV}&=&(1.53\pm0.10)\times10^{-15}~{\rm TeV^{-1}},\label{slope1}\\
 \Delta t_{\rm in} &=&(1.7\pm3.6)\times10^3~{\rm s}.\label{intercept1}
  \end{eqnarray}
This combined linear fitting result is well consistent with the TeV regularity equations~(\ref{slope}) and (\ref{intercept}).
The LV scale is obtained immediately,
 \begin{equation}
 E_{\rm LV}=(6.5\pm0.4)\times 10^{17}~\rm{GeV}.\label{LV scale}
\end{equation}
From previous GBR photon studies~\cite{Zhang:2014wpb,Xu:2016zsa,Xu:2016zxi,Amelino-Camelia:2016ohi,Xu:2018ien}, the LV scale was determined as $E_{\rm LV}=3.6\times10^{17}~{\rm GeV}$. Considering the gap over five orders of magnitude in energy scale, the LV scale obtained by the neutrino data is essentially in agreement with the GRB photon result. Such a consistency between different particles and energy scales can be considered as a positive support for the energy dependent speed variation of ultra-relativistic particles. For the GRB photons, there still exist alternative results, such as the one from an analysis of the short burst GRB 090510~\cite{Ackermann:2009aa} and also analyses with different data selection criteria as have been discussed in previous works~\cite{Amelino-Camelia:2016ohi,Xu:2018ien}. Therefore the conclusion on the light-speed variation still needs to be tested by more studies in future.

\subsection*{Lorentz violation as an explanation}

The similar regularity of energy dependent speed variation is supported by the analysis on GRB photons and GRB neutrinos with different energies. Theoretically, the speed variation may be caused by different reasons. The \textit{in vacuo} dispersion due to LV is one of the probable options, as well as the matter effect. The cosmic matters on the path might alter the propagation time of particles, but for the ultra-relativistic particles, the matter effect can only cause the time delay, i.e., $\Delta t>0$. However, both time advance and time delay cases exist for either TeV and PeV neutrinos, as revealed by previous studies~\cite{Amelino-Camelia:2016fuh,Amelino-Camelia:2016ohi} and our analysis. Different signs of time correction can not be interpreted by the pure matter effect. Because of the CPT odd feature of the linear Lorentz correction,  neutrinos and anti-neutrinos have  different signs for the LV sign factor $s$~\cite{Stecker:2014oxa}, thus they can be superluminal and subluminal respectively (or vice versa) in propagation. So the Lorentz invariant violation effect is a more reasonable explanation. We therefore reveal that neutrinos and anti-neutrinos have different properties to break the charge, parity and time (CPT) reversal symmetry.

\section*{Discussion}

Inferences based on ultra-high energy neutrinos could overcome several interference factors. To obtain stronger limits and higher time resolution, particles are expected to come from more distant sources. The propagation distance of high-energy photons is limited by the pair production, so GRB neutrinos have more advantages over photons to probe LV physics~\cite{Jacob:2006gn}. On the other hand, ultra-high energy neutrinos can be distinguished from neutrinos produced in the atmosphere and other backgrounds. As mentioned in refs~\cite{Aartsen:2016qcr,Aartsen:2017wea}, the detection results of TeV neutrino events are still found consistent with backgrounds. However, ultra-high energy neutrinos above 1~PeV stand out from backgrounds, thus render the revealed regularity more convincing. In addition, little information about intrinsic time difference was known until now. If $\Delta t_{\rm obs}$ is not too long, $\Delta t_{\rm in}$ might have a significant impact on the results. But compared with the LV time correction $\Delta t_{\rm LV}$ of up to months for PeV neutrinos, $\Delta t_{\rm in}$ on the order of 1-2 hours can be safely neglected in our analysis.

In summary, we propose for the first time the association of IceCube PeV events with GRB candidates. By analysing the TeV and PeV IceCube neutrino events that likely associated with GRBs, we find that the GRB neutrino events fall on a pair of lines in the $\Delta t_{\rm obs}/(1+z)$ versus $K$ plot. All four PeV neutrinos detected so far  agree with the regularity revealed from TeV neutrinos, implying an energy dependent speed variation of ultra-high energy neutrinos. This regularity is similar to the light speed variation previously proposed from GRB photons. The LV scale $E_{\rm LV}=(6.5\pm 0.4)\times10^{17}~{\rm GeV}$ determined by GRB neutrino events is comparable with that determined by GRB photons. We also suggest that neutrinos and anti-neutrinos have different signs of LV time correction $\Delta t_{\rm LV}$, so that there are both ``early'' and ``late'' events. Certainly, these results still remain to be tested by more data in the future. Since the IceCube Collaboration as well as many other researchers are advancing the coincidence with ultra-high energy neutrinos and GRBs~(see e.g. ref.~\cite{Aartsen:2017wea} and references therein), it is expected that the energy dependent speed variation accompanied with other Lorentz violation features can be tested in the foreseeable future.

\section*{Methods}
\footnotesize
\subsection*{Estimation of error range.}

The errors in our analysis are based on the uncertainties of energy, redshift and cosmological parameters. Since the errors of cosmological parameters are provided by PDG~\cite{Agashe:2014kda}, we focus on the energy and redshift here. For the energy, we can get both positive and negative errors from the IceCube database, but the positive errors are higher in our estimation. Although the IceCube detector has very good energy resolution, the detector may not collect the entire energy of particles produced by ultra-high energy neutrinos, since the interaction vertex of neutrino events may be located outside the instrumental volume. The energy $E$ provided by the IceCube Collaboration is regarded as an approximate lower limit of the neutrino energy~\cite{Aartsen:2014aqy}. Furthermore, the uncertainties are different between ``shower'' and ``track'' events. The ``deposited energy'' (the energy information given by IceCube databases) is close to the neutrino energy for most shower events, but for track events the deposited energy is only a lower bound to the true neutrino energy. Among our four PeV events, three of them are shower events, while the fourth is a track event. So in our analysis, the positive error of energy is set as $30\%$ for three shower events and $50\%$ for the fourth track event as a reasonable ``explorative assumption'', and the negative error is still provided by IceCube measurements. In practice, different energy error estimations of the fourth track event do not make significant changes in fitting results. Even if we assume that the deposited energy of the track event is only half of the true energy and do the same linear fitting, we find that the new slope $(1.36\pm0.13)\times10^{-15}~{\rm TeV^{-1}}$ is consistent with the slope $(1.53\pm0.10)\times10^{-15}~{\rm TeV^{-1}}$ in Fig.~\ref{fig4}, if taking into account uncertainties. The new correlation coefficient $r=0.95$ also implies a relatively strong linear correlation.

For the redshift, some of our GRB candidates do not have a determined redshift yet. The likely estimated value is obtained as the average of known redshifts in previous analyses~\cite{Amelino-Camelia:2016fuh,Amelino-Camelia:2016ohi}. Here we follow this principle and estimate $z=2.15$ for ``long bursts'' and $z=0.5$ for ``short bursts'', as suggested in the IceCube database~\cite{IceCube website}. As an average of all ``long burst'' redshifts that have been determined so far, $z=2.15$ also corresponds closely to the average of two known redshift values of our PeV GRB neutrino candidates. The error of the known redshift is negligible, since the measurement of the redshift is extraordinarily accurate currently.  The likely error range of the unknown redshift is suggested as $0.5z\sim2z$ in our analysis. This estimation method can also separate the two kinds of GRBs.

\begin{table*}[h]
	\centering
	\caption{\sf \small The properties of PeV neutrino events with associated GRB candidates. }
	
	\label{tab:3}
	
	\begin{tabular}{llllllll}
		
		\hline\hline
		\    \   & $E$~(PeV)  & $\sigma$ & $\Delta\Psi$ & $z$ & $\Delta t_{\rm obs}~(10^3\rm s)$ & $\frac{\Delta t_{\rm obs}}{1+z}~(10^3\rm s)$ & $K~(10^{18}{\rm s\cdot TeV})$ \    \   \\
		\hline
		\noalign{\vspace{0.5ex}}
		\noalign{\vspace{0.5ex}}
		\    \  event $\#14$ & $1.04_{-0.14}^{+0.13}$ &  $13.2^\circ$ & &  &  &  &    \    \   \\
		\noalign{\vspace{0.5ex}}
		\    \   GRB 110725A$^\dagger$& & $9.06^\circ$ & $4.87^\circ$ & $2.15^*$ & 1320.217 & 419.1 & 350.2 \    \   \\
		\noalign{\vspace{0.5ex}}
		\    \   GRB 110730A$^\natural$& & $4.28^\circ$ & $5.6^\circ$ & $2.15^*$ & 907.885 & 288.2 & 350.2 \    \   \\
		\noalign{\vspace{0.5ex}}
		\    \   GRB 110731A& & $0.0001^\circ$ & $13.14^\circ$ & $2.83$ & 782.096 & 204.2 & 366.9 \    \   \\
		\noalign{\vspace{0.5ex}}
		\    \   GRB 110808B& & $0.0693^\circ$ & $9.8^\circ$ & $0.5^*$ & 74.303 & 49.5 & 172.8 \    \   \\
		\noalign{\vspace{0.5ex}}
		\    \   GRB 110905A& & $0.0314^\circ$ & $14.9^\circ$ & $2.15^*$ & -2309.121 & -733.1& 350.2 \    \   \\
		\noalign{\vspace{0.5ex}}
		
		\hline
		\noalign{\vspace{0.5ex}}
		\noalign{\vspace{0.5ex}}
		\    \  event $\#20$ & $1.14_{-0.138}^{+0.14}$ &  $10.7^\circ$ & &  &  &  &    \    \   \\
		\noalign{\vspace{0.5ex}}
		\    \   GRB  111229A$^\natural$& & $0.0003^\circ$ & $18.9^\circ$ & 1.3805 & 384.970 & 161.7 & 355.4 \    \   \\
		\noalign{\vspace{0.5ex}}
		\    \   GRB  120119C$^\dagger$& & $4.42^\circ$ & $36.9^\circ$ & $2.15^*$ & -1940.176 & -615.9 & 383.9 \    \   \\
		\noalign{\vspace{0.5ex}}
		\    \   GRB  120210A& & $5.51^\circ$ & $11.4^\circ$ & $0.5^*$ & -3304.901 & -2203.3 & 189.4 \    \   \\
		\noalign{\vspace{0.5ex}}
		\hline
		\noalign{\vspace{0.5ex}}
		\noalign{\vspace{0.5ex}}
		\    \  event $\#35$ & $2.00_{-0.26}^{+0.24}$ &  $15.9^\circ$ & &  &  &  &    \    \   \\
		\noalign{\vspace{0.5ex}}
		\    \   GRB  120919A& & $0.0863^\circ$ & $11.0^\circ$ & $2.15^*$ & 6539.722 & 2076.1 & 674.3 \    \   \\
		\noalign{\vspace{0.5ex}}
		\    \   GRB  121229A$^\natural$& & $0.0003^\circ$ & $12.1^\circ$ & 2.707 & -2091.621 & -564.2 & 702.5 \    \   \\
		\noalign{\vspace{0.5ex}}
		\    \   GRB  130121A$^\dagger$& & $1.14^\circ$ & $6.55^\circ$ & $2.15^*$ & -4046.519 & -1284.6 & 674.3 \    \   \\
		\noalign{\vspace{0.5ex}}
		\hline
		\noalign{\vspace{0.5ex}}
		\noalign{\vspace{0.5ex}}
		\    \  ATel $\#7856$ & $2.6_{-0.3}^{+0.3}$ &  $1^\circ$ & &  &  &  &    \    \   \\
		\noalign{\vspace{0.5ex}}
		\    \   GRB  140427A$^\dagger$& & $23.26^\circ$ & $25.8^\circ$ & $2.15^*$ & 3827.439 & 1215.1 & 874.9 \    \   \\
		\noalign{\vspace{0.5ex}}
		\    \   GRB  140516B$^\natural$& & $7.77^\circ$ & $8.63^\circ$ & $2.15^*$ & 2185.942 & 693.9 & 874.9 \    \   \\
		\noalign{\vspace{0.5ex}}
		\hline\hline
		
		\noalign{\vspace{0.5ex}}
	\end{tabular}
	
	\footnotesize
	
	The energy errors here are measurement uncertainties provided by the IceCube database. The column $\sigma$ shows angular uncertainties of neutrino events and GRB candidates respectively. The angular separation $\Delta \Psi$ is calculated from the differences between RA and Dec angles. For every one of the four events, there exists a candidate marked by $^\dag$ that satisfies the strict time criterion and is consistent with the regularity of the TeV neutrino. The mark $^\natural$ represents another option with a strong correlation.
\end{table*}

\subsection*{Time and direction criteria.}

We adopt time and direction criteria to select the associated GRBs of our PeV neutrino events from the GRBs database on the IceCube web interface. 
The time correction of the LV effect could be extended by the ultra-high energy and the long propagation distance of neutrinos. So it is reasonable to expand the time range with the increase of energy. We include the GRBs detected within one month before or after the neutrino for the two 1~PeV events, within two months for the 2~PeV event, and within three months for the 2.6~PeV event.

We also require that the associated GRB has a consistent direction with the neutrino event. As the directional criterion, a two dimensional circular Gaussian~\cite{Amelino-Camelia:2016fuh}
\begin{equation}
P(\nu, \mathrm{GRB})=\frac{1}{2\pi\sigma^2}\exp(-\frac{\Delta \Psi^2}{2\sigma^2}),
\end{equation}
is introduced, where $\Delta \Psi$ is the angular separation between GRB and neutrino, and $\sigma=\sqrt{\sigma_{\rm GRB}^2+\sigma_{\nu}^2}$ is the standard deviation based on the angular uncertainties of GRB and neutrino measurements. In our analysis, we consider the GRBs whose angular separation is smaller than $3\sigma$. The properties of GRB candidates that satisfy both time and direction criteria are listed in Table~\ref{tab:3}.

To test the consistency between TeV and PeV neutrinos, we use a strict time criterion that requires
\begin{equation}
|\frac{\Delta t_{\rm obs}}{1+z}-s\cdot\frac{K}{E'_{\rm LV}}|<30\%\cdot\frac{K}{E'_{\rm LV}}\label{strict time},
\end{equation}
where $E'_{\rm LV}$ is obtained from equation~(\ref{slope}) and represents the regularity of the TeV neutrinos. All four PeV events have associated GRBs satisfing the strict time criterion. We mark these suggested GRBs with $^\dag$ in Table~\ref{tab:2} and Table~\ref{tab:3}, and include them in the new linear fitting in Fig~\ref{fig4}. For the event $\#14$, both GRB 110725A and GRB 110905A satisfy equation~(\ref{strict time}), and we select GRB 110725A according to the maximal correlation criterion~\cite{Amelino-Camelia:2016fuh,Amelino-Camelia:2016ohi}, which requires the correlation coefficient $r$ to be of maximal value in the fitting of the selected GRBs. In fact, the results of linear fittings with either GRB 110725A or GRB 110905A are in agreement within the error range.
The angular separation of the suggested GRB 120118C with the IceCube event $\#20$ is about $3\sigma$.
The other three of the four suggested GRBs, i.e., GRB 110725A, GRB 130121A and GRB 140427A, have an angular separation $\Delta \Psi$ less than or close to $1\sigma$, which represents an extraordinary direction consistency to render a convincing regularity.

Since each PeV neutrino event has multiple GRB candidates, there is more than one option to combine candidates as a group. Solely considering the four PeV events, we can calculate the correlation coefficient of all options and pick the group with maximal correlation. Coincidentally, it is the suggested group marked by $^\dag$ that gives the highest value of correlation, which is $r=0.996$. Therefore our suggested GRB candidates selected by equation~(\ref{strict time}) satisfy the maximal correlation criterion mentioned in refs~\cite{Amelino-Camelia:2016ohi,Amelino-Camelia:2016fuh} as well. It is noteworthy that another option marked by $^\natural$ in Table~\ref{tab:3} also has a strong correlation and gives $r=0.994$ with $E_{\rm LV}=(12.7\pm 0.1)\times10^{17}~{\rm GeV}$. But in the linear fitting of both TeV and PeV events, we find that this option results in a large negative intrinsic time difference $\Delta t_{\rm in}=(-72.9\pm0.3)\times10^3~{\rm s}$, which leads to an unnatural vision that the ultra-high energy neutrinos are emitted about 20 hours before the GRB photons. So the $^\natural$ marked candidates are less favored as the $^\dag$ marked ones suggested by us.

The association between GRBs and PeV neutrinos was rarely mentioned in previous studies, since people focused on looking for associated GRBs with very close temporal coincidence with neutrino events. Only a blazar PKS B1424-418 was proposed to be associated to the PeV event $\#35$~\cite{Kadler:2016ygj} because the neutrino event was detected on 4 December 2012,
within the PKS B1424-418 outburst period from 16 July 2012 to 30 April 2013. In this work, we extend the time range and find that all four events of PeV neutrinos detected so far might be associated to the GRB candidates listed in Table~\ref{tab:3}, if taking into account the LV effect. For the event $\#35$, the suggested GRB 130121A was detected during the PKS B1424-418 outburst period. The angular separation of GRB 130121A $\Delta \Psi=6.55^\circ \approx 0.4 \sigma$ is smaller than that of PKS B1424-418. The three probable GRB candidates of event $\#35$ indicate a range of LV scale, which is $(3.25-12.45)\times10^{17}~\rm GeV$, whereas the limit from the association between event $\#35$ and PKS B1424-418 gives
a linear LV scale $E_{\rm LV}> 0.01E_{\rm Pl}\approx1.22\times10^{17}~{\rm GeV}$~\cite{Wang:2016lne}. So even if the event $\#35$ was emitted by the blazar PKS B1424-418 instead of GRB 130121A, it is still compatible with our result of $E_{\rm LV}=(6.5\pm 0.4)\times10^{17}~{\rm GeV}$.

\subsection*{Data availability.}
The data that support the plots within this paper and other
findings of this study are available from the corresponding author upon
reasonable request.


\section*{Acknowledgements}
\footnotesize
 This work is supported by National Natural Science Foundation of China (Grant No.~11475006).

\section*{Author contributions}
\footnotesize
Both authors contributed extensively to the work presented in this paper.

\section*{Additional information}
\footnotesize
 \subsection*{Competing interests:}
The authors declare no competing Interests.

\end{document}